\documentclass[
	a4paper, 
	10pt,    
	twoside, 
]{LTJournalArticle}

\usepackage{amssymb}
\usepackage{amsmath}

\newcommand{\aap}{Astron. Astrophys.}
\newcommand{\aaps}{Astron. Astrophys. Suppl. Ser.}
\newcommand{\apj}{Astrophys. J.}
\newcommand{\apjl}{Astrophys. J. Lett.}
\newcommand{\apjs}{Astrophys. J. Suppl. Ser.}

\newcommand{\mnras}{MNRAS}
\newcommand{\nat}{Nature}

\newcommand{\pra}{Physical Review A}

\newcommand{\ssr}{Space Science Reviews}

\runninghead{Passage of a GRB Through a Molecular Cloud} 

\footertext{\textit{Astronomy Letters} (2024) 50:99-119} 

\title{Passage of a Gamma-Ray Burst Through a Molecular Cloud:\\ Cloud Ionization Structure} 

\author{%
	A.V. Nesterenok\textsuperscript{1}\thanks{Corresponding author: \href{mailto:alex-n10@yandex.ru}{alex-n10@yandex.ru}\\ \textbf{Received:} December 4, 2023, \textbf{Published:} Febrary, 2024}
}

\date{\footnotesize\textsuperscript{\textbf{1}}Ioffe Institute, Saint Petersburg, Russia}

\renewcommand{\maketitlehookd}{%
	\begin{abstract}
	\normalsize \rm
		\noindent The model is constructed of the propagation of gamma-ray burst radiation through a dense molecular cloud. The main processes of the interaction of the radiation with the interstellar gas are taken into account in the simulations: the ionization of H and He atoms, the ionization of metal ions and the emission of Auger electrons, the photoionization and the photodissociation of H$_2$ molecules, the absorption of ultraviolet radiation via Lyman and Werner band transitions of H$_2$, the thermal sublimation of dust grains. The ionization of metal ions by X-ray radiation determines the gas ionization fraction in the cloud region, where the gas is predominantly neutral. The ionization of inner electron shells of ions is accompanied by the emission of Auger electrons, giving rise to metal ions in a high ionization state. In particular, the column densities of Mg, Si, Fe in the ionization states I--IV are much lower than the column densities of these ions in the ionization state V and higher. The photoionization of metal ions by ultraviolet radiation takes place at distances smaller than the dust-destruction radius and only for neutral atoms with an ionization threshold below 13.6~eV. The simulations have confirmed the previously made suggestion that the ionization of He atoms plays an important role in the absorption of radiation in the X-ray wavelength range. For a low metallicity, $\rm [M/H] \leq -1$, the role of He atoms is dominant. 		
	\end{abstract}
}

\begin{document}
\maketitle 

\section{Introduction}
A distinctive feature of gamma-ray bursts is the release into the interstellar medium of a huge amount of ionizing radiation within a few tens of seconds. Gamma-ray bursts are divided into two types with different durations of gamma-ray radiation: short ($\leq 1-2$~s) and long ($\geq 1-2$~s) \autocite{Pozanenko2021}. 
Until recently the long gamma-ray bursts have been thought to be produced by the collapse of massive stars. However, the observational data have begun to emerge recently that give evidence that the long gamma-ray bursts are produced both as a result of massive star collapse and during the compact star mergers (e.g. \autocite{Rastinejad2022,Petrosian2024}).
The energy of the explosion that results in the gamma-ray burst can reach $10^{52}$~erg and higher \autocite{Goldstein2016}. Long gamma-ray bursts are detected both in galaxies of the Local Universe and in distant galaxies at redshifts up to $z \approx 9$ \autocite{Cucchiara2011}.
Owing to the high intensity and wide frequency range of the gamma-ray burst emission (the prompt emission and the afterglow of the gamma-ray burst extend from radio to gamma-ray wavelength ranges), the gamma-ray bursts are probes of the interstellar medium of their host galaxies and absorption systems on the line of sight \autocite{Schady2017}. 

The generally accepted model of the gamma-ray bursts is the "fireball" model \autocite{Kumar2015}. According to this model, the compact "central engine" that is produced by the collapse of a massive star or by the compact star merger, launches a relativistic plasma jet.
There are two main stages in the formation of the gamma-ray burst radiation. The first stage of the radiation formation is associated with processes inside the relativistic jet (the photospheric origin of the radiation, the energy dissipation in shocks).
At this stage the prompt emission of the burst is produced. The emission of this relatively short prompt phase lies in the X-ray and gamma-ray wavelength ranges. The second stage includes the interaction of the jet with the external circumburst medium, and the formation of forward and reverse shocks \autocite{Sari1999}.
At this stage an optical flash (the emission of the reverse shock) and the afterglow (the emission of the forward shock) are produced by the synchrotron mechanism. The optical flash (or "prompt" optical emission) is detected for a small number of bursts. The characteristic time of the optical flash is $\sim 10^2-10^3$~s \autocite{Oganesyan2023}.
This emission coincides in time or immediately follows the prompt gamma-ray phase. While the afterglow of gamma-ray burst can be detected for several weeks in the optical and X-ray wavelength ranges, and for several months or more -- in the radio band. For some gamma-ray bursts, the emission of the associated type Ic supernova is observed \autocite{Pozanenko2021}.

An analysis of the absorption lines in the afterglows of gamma-ray bursts allows one to investigate the parameters of the interstellar medium in the close proximity to the burst progenitor, the parameters of gas--dust clouds inside the host galaxy and in the galactic halo (e.g., \autocite{Prochaska2008,Fox2008,Heintz2018}). Observations of the hydrogen line ${\rm Ly} \alpha$ with optical ground-based telescopes are possible for gamma-ray bursts at redshifts $z \geq 1.6$.
Most of the gamma-ray bursts at such redshifts have ${\rm Ly} \alpha$ absorption lines in the afterglow spectra. The measured values of hydrogen column densities $N_{\rm HI}$ reach up to a few $10^{22}$~cm$^{-2}$ for some gamma-ray bursts, and the median value of $N_{\rm HI}$ is about $4 \times 10^{21}$~cm$^{-2}$ \autocite{Fynbo2009,Schady2017,Tanvir2019}.  
There is a large scatter of measured $N_{\rm HI}$ values at all redshifts $z \geq 1.6$. But at the same time the measured values of $N_{\rm HI}$ are higher on average for gamma-ray bursts than the observed values of $N_{\rm HI}$ for absorption systems in the spectra of quasars. This may be a consequence of the fact that gamma-ray bursts preferentially trace the dense environment within host galaxies, where the star-formation processes proceed \autocite{Prochaska2007,Thone2013}. 
In the infrared and optical wavelength ranges, the spectral energy distribution of the gamma-ray burst afterglow is fitted by taking into account the interstellar absorption in the Milky Way and in the host galaxy. This allows one to determine visual extinction $A_{\rm V}$ in the host galaxy and, on this basis, to estimate the hydrogen column density $N_{\rm H}$.

The hydrogen column density $N_{\rm HX}$ can be determined based on the absorption of the gamma-ray burst afterglow in the X-ray part of the spectrum. It is usually assumed in these estimates that the radiation is absorbed by metal ions. The column densities $N_{\rm HX}$ were found to be, as a rule, an order of magnitude or more higher than hydrogen column densities $N_{\rm H}$, calculated based on the data in the optical wavelength range -- based on absorption lines of atomic hydrogen and metal ions, or $A_{\rm V}$ \autocite{Watson2007,Campana2010,Campana2012}. 
One of the solutions of the discrepancy between the observed  hydrogen column densities is that the afterglow radiation in the X-ray wavelength range is absorbed by a layer of ionized gas in the vicinity of the gamma-ray burst source \autocite{Schady2011,Krongold2013}. The metal ions with a high charge contribute to the afterglow absorption in the X-ray wavelength range, but are not visible in the optical and ultraviolet (UV) part of the spectrum.
This explanation is supported by the fact that the analysis of the absorption-line variability of low-charge metal ions in the afterglow of gamma-ray bursts provides that clouds of neutral gas are at a distance of $\sim 100-1000$~pc from the burst progenitor (e.g., \autocite{Vreeswijk2007}).
Other explanations for the observed absorption of afterglow radiation in the X-ray wavelength range have been suggested -- the absorption by helium ions in the HII region, where the gamma-ray burst occurred \autocite{Watson2013}. The diffuse intergalactic medium and intervening absorbing clouds along the line of sight contribute significantly to the absorption of radiation for gamma-ray bursts at redshifts $z \geq 3$ \autocite{Starling2013,Rahin2019}.  

Massive stars have a short lifetime, $t \lesssim 10^7$~years, and, therefore, the explosion of a massive star occurs with a high probability in the region of its birth \autocite{Crowther2012}. The star-forming regions are characterized by relatively high densities of interstellar gas. Therefore, it is possible that the radiation of the gamma-ray burst passes through a dense molecular cloud.
A number of studies have been devoted to the interaction of intense gamma-ray burst radiation with the interstellar gas near the burst progenitor star (e.g., \autocite{Draine2002,Perna2002,Perna2003,Lazzati2003,Barkov2005, Barkov2005b,Badin2010}). Each of these papers considered one aspect of the problem.
In particular, \textcite{Barkov2005,Barkov2005b} considered the effect of gamma-ray burst radiation on the molecular gas at a distance up to 1.5~pc from the burst source. There is a full gas ionization by X-ray and gamma-ray radiation at such distances, and the gas is heated to high temperatures, $T_{\rm g} \gtrsim 10^4$~K. 
In our paper we examine the effect of gamma-ray burst radiation on the molecular gas at larger distances from the burst progenitor star, several parsecs or more. The numerical method that is used in our paper is analogous to the approach by \textcite{Draine2002}. \textcite{Draine2002} considered the UV radiation of the gamma-ray burst (optical flash) in their model, and they did not take into account the ionization of metal ions.  
The UV and X-ray radiation of the gamma-ray burst is taken into account in our model. It is shown that the ionization of metal ions should be taken into consideration in the simulations of the ionization structure of the cloud. The aim of this work is to determine the components of the molecular gas (H$_2$, He, metal ions, dust) that make a major contribution to the absorption of gamma-ray burst afterglow emission in the optical and X-ray wavelength ranges.

\section{The ionization and dissociation processes}
\subsection{The photoionization processes}
The main channel of H$_2$ photoionization is the reaction:

\begin{equation}
H_2 + h\nu \to H_2^+ + e^-.
\label{eq_h2_ion}
\end{equation}

\noindent
The analytical expression for the cross section of this reaction is given by \autocite{Yan1998,Yan2001}. The ratio of cross sections for photoionization of H$_2$ and H is 2.8 at high energies. Another photoionization channel of H$_2$ is ionization accompanied by dissociation:

\begin{equation}
H_2 + h\nu \to H^+ + H + e^-.
\label{eq_h2_dission}
\end{equation}

\noindent
The experimental data on the cross section of the reaction (\ref{eq_h2_dission}) for photon energies $18-124$~eV are given in the paper by \textcite{Chung1993}. The measured cross section includes the contribution of double photoionization of H$_2$, which is about 20 per cent at photon energies 80~eV. The double photoionization of H$_2$ is neglected in our calculations. The dissociative photoionization cross section is about $25-30$ per cent of the cross section of the reaction (\ref{eq_h2_ion}). We neglect the dependence of H$_2$ ionization cross sections on the rotational--vibrational state of the molecule in calculations of ionization rates.

The accurate quantum mechanical expression is used for the photoionization cross sections of H atoms and HeII ions \autocite{Osterbrock2006}. We use the analytic approximations for the photoionization cross sections of HeI and metal ions published by \textcite{Verner1995,Verner1996}. The ionization of the inner shells of metal ions is accompanied by the ejection of Auger electrons. The probability distribution for the number of ejected electrons as a result of the Auger effect are taken from \textcite{Kaastra1993}.

The photodissociation and photoionization of the ion H$_2^+$ are taken into account in the model:

\begin{equation}
\begin{split}
& H_2^+ + h\nu \to H^+ + H, \\
& H_2^+ + h\nu \to H^+ + H^+ + e^-.
\end{split}
\end{equation}

\noindent
For the photodissociation of H$_2^+$, we use the cross section approximation from \textcite{Draine2002}, which was derived from the experimental data published by \textcite{vonBusch1972}. The photodissociation cross section is averaged over the H$_2^+$ vibrational distribution. The cross section has no energy threshold, therefore this process contributes to the absorption of radiation in the optical and UV wavelength ranges \autocite{Perna2003}. The cross section calculated by \textcite{Arkhipov2018} is used for the H$_2^+$ photoionization.

Note that the analytic expression for the cross section of HeI photoionization provided by \textcite{Draine2002}, is larger than the cross section from the paper by \textcite{Verner1996} by more than an order of magnitude at photon energies $E \gtrsim 1$~keV. 

\subsection{The Compton ionization}
The cross sections for the Compton scattering by electrons bound in H and He atoms, H$_2$ molecules, are taken from \textcite{Hubbell1975}. These cross sections approach the Klein-Nishina-Tamm cross section at photon energies $E > 3-5$~keV. It is at such photon energies that the Compton ionization becomes the main ionization process of H, He, H$_2$. The cross sections for the Compton ionization of metal ions are calculated based on the formula for the cross section of the Compton scattering by free electrons. 

\subsection{The H$_2$ photodissociation via absorption in Lyman and Werner bands}
The radiation in the UV wavelength range is absorbed by H$_2$ molecules via Lyman and Werner band transitions. The decay time of electronically excited states $B^1\Sigma^+_u$ and $C^1\Pi^{\pm}_u$ of H$_2$ molecule is $\sim 10^{-9}$~s. And the excited H$_2$ molecule will rapidly decay back to the ground electronic state:

\begin{equation}
\begin{gathered}
H_2(X,v'',J'') + h\nu \to H^{*}_2(B,C;v',J') \to \\
\to H_2(X,v,J) + h\nu'.
\end{gathered}
\end{equation}

\noindent
The process is possible when the excited H$_2$ molecule decay by a
transition to the vibrational continuum of the ground electronic state -- this H$_2$ dissociation mechanism is known as the Solomon process:

\begin{equation}
\begin{gathered}
H_2(X,v'',J'') + h\nu \to H^{*}_2(B,C;v',J') \to \\
\to H(1s) + H(1s) + h\nu'.
\end{gathered}
\end{equation}
 
\noindent
The fraction of H$_2$ excitations that bring to the dissociation is about 15 per cent on average \autocite{Draine1996}.

The direct continuum photodissociation of H$_2$ molecule results from the transition from the ground electronic state $X^1\Sigma^{+}_{g}$ into the ro-vibrational continuum of excited electronic states,

\begin{equation}
H_2 (X, v'', J'') + h\nu \to H(1s) + H(2p).
\label{eq_h2_diss_cont}
\end{equation}

\noindent
\textcite{Gay2012} performed cross section calculations for the process (\ref{eq_h2_diss_cont}) for the transitions from $(v'',J'')$ levels of the ground electronic state to the ro-vibrational continuum of $B^1\Sigma^{+}_{u}$ and $C^{1}\Pi_{u}$ states.

\begin{table*} 
	\caption{\normalsize Parameters of the gamma-ray burst emission.}
	\centering 
	\begin{tabular}{L{0.41\linewidth} L{0.2\linewidth}} 
	\toprule
	\multicolumn{2}{c}{Prompt emission} \\
	Isotropic equivalent energy, $E_{\gamma, {\rm iso}}$ & $5 \times 10^{52}$~erg \\
	Peak energy, $E_{\rm peak}$ & 350~keV \\
	$\alpha$ & $-1$ \\
    $\beta$ & $-2.3$ \\
	\midrule
	\multicolumn{2}{c}{Afterglow} \\
	Kinetic energy of the blast-wave, $E_{\rm K}$ & $2.5 \times 10^{53}$~erg \\
	Fraction of energy in the magnetic field, $\varepsilon_{\rm B}$ & $10^{-4}$ \\
	Fraction of energy in electrons, $\varepsilon_{\rm e}$ & 0.1 \\
	Electron spectral index, $p$ & 2.3 \\
	Jet half-opening angle, $\theta_{\rm j}$ & 0.1~rad \\
	Circumburst medium density, $n_{0}$ & 1~cm$^{-3}$ \\
	\midrule
	\multicolumn{2}{c}{Optical flash} \\
	Peak isotropic equivalent luminosity, $L_{0}$ & $3 \times 10^{32}$~erg~s$^{-1}$~Hz$^{-1}$ \\
	Time when the luminosity peak is reached, $t_0$ & 10~s \\
	\bottomrule
	\end{tabular}
\end{table*}

\section{Description of the numerical model}
\subsection{The intensity of gamma-ray burst radiation}
The prompt emission of the gamma-ray burst is described by two main parameters: the total isotropic equivalent radiated energy $E_{\gamma, {\rm iso}}$ and the peak energy $E_{\rm peak}$ of the spectral energy distribution $\nu F_{\nu}$ in the rest frame of the gamma-ray burst.
The parameter $E_{\gamma, {\rm iso}}$ is the total energy of prompt emission of the gamma-ray burst in the rest-frame energy range $1-10^4$~keV assuming an isotropic distribution of the emission. The correlation between $E_{\gamma, {\rm iso}}$ and $E_{\rm peak}$ is called the Amati relation \autocite{Amati2006}. In our model we take $E_{\gamma, {\rm iso}} = 5 \times 10^{52}$~erg and $E_{\rm peak} = 350$~keV \autocite{Nava2012, Tsvetkova2021}. 
The flux density spectrum of the prompt emission is approximated by the Band function \autocite{Band1993}. We take the following values for the spectral indices for the power-law spectrum at low and high energies:  $\alpha = -1$ and $\beta = -2.3$, respectively \autocite{Kaneko2006,Nava2011}. The time dependence of the radiation intensity consists of an increasing linear function ($0 \leq t \leq t_0 = $ 1~s) and an exponential decay with a characteristic time $t_{\rm exp} = 5$~s.

The isotropic kinetic energy in the gamma-ray burst blast-wave $E_{\rm K}$ can be inferred indirectly from the observed flux and duration of the afterglow. The sum of $E_{\gamma, {\rm iso}}$ and $E_{\rm K}$ gives the total amount of initial explosion energy. The efficiency of the prompt phase of the gamma-ray burst $\eta$ is defined as the ratio of the energy radiated in the prompt phase $E_{\gamma, {\rm iso}}$ and the total explosion energy.
The estimates of the parameter $\eta$ based on the gamma-ray burst observations lie in a wide range, from $\eta \lesssim 0.1$ to $\eta \approx 0.9$ \autocite{Zhang2007,Beniamini2016}. In our calculations we take $E_{\rm K} = 2.5 \times 10^{53}$~erg and $\eta = 0.17$. The flux density spectrum of the afterglow and its evolution with time is calculated with the {\it afterglowpy} code \autocite{Ryan2020}.
The following parameters are used in the calculations of the radiation intensity of the gamma-ray burst afterglow: the fraction of the forward shock energy that is in the magnetic field $\varepsilon_{\rm B} = 10^{-4}$ \autocite{Santana2014,BarniolDuran2014}, the fraction of energy that resides in electrons $\varepsilon_{\rm e} = 0.1$ \autocite{Nava2014,Beniamini2017}, the electron spectral index $p = 2.3$ \autocite{Curran2010,Ryan2015}, the jet half-opening angle $\theta_{\rm j} = 0.1$~rad \autocite{Ryan2015,Goldstein2016}.
The density of the circumburst medium is considered to be constant and equals to $n_{0} = 1$~cm$^{-3}$ \autocite{Panaitescu2002}. The observer is on the jet axis. Our afterglow emission model in {\it afterglowpy} does not include the energy injection into the blast-wave. 

The optical flash is observed simultaneously or immediately after the prompt emission phase of the gamma-ray burst \autocite{Oganesyan2023}. One of the possible mechanisms of the optical flash is the synchrotron radiation of the reverse shock, and it is generated at the initial stage of the jet propagation in the circumburst medium. In our model, the flux of the optical flash is approximated by the following function of time \autocite{Nakar2004}:

\begin{equation}
F_{\rm opt} = F_{0, {\rm opt}} \left[ \frac{1}{2} \left(\frac{t}{t_0}\right)^{-s\alpha_1} + \frac{1}{2} \left(\frac{t}{t_0}\right)^{-s\alpha_2}\right]^{-(1/s)},
\end{equation} 

\noindent
where the parameters $\alpha_1 = 0.5$, $\alpha_2 = -2$, $s = 2$, $t_0$ is the time when the radiation flux reaches its maximum, $F_{0, {\rm opt}}$ is the flux at the time $t_0$. In our model $t_0 = 10$~s (in the burst rest frame). To estimate the isotropic equivalent luminosity of the optical flash we use the results by \textcite{Nakar2004,Nakar2005}. For the parameters presented in Table I we have: 

\begin{equation}
L_{0} = 4 \pi D_{\rm L}^2 F_{0, {\rm opt}} \approx 3 \times 10^{32}~\textrm{erg}~\textrm{s}^{-1}~\textrm{Hz}^{-1},
\label{eq_opt_lum_ampl}
\end{equation}

\noindent
where $D_{\rm L}$ is the luminosity distance from the gamma-ray burst progenitor to the observer. The frequency dependence of the luminosity is assumed to be as follows:

\begin{equation}
L(\nu) = L_{0} \left(\frac{\nu_0}{\nu}\right)^{\beta},
\end{equation}

\noindent
where $\nu_0 = 5.4 \times 10^{14}$~Hz, $\beta = 1$. The value of $L_{0}$ considered in our model is a factor of $\approx 500$ lower than the luminosities of the optical flash of the bright gamma-ray bursts GRB~990123 \autocite{Akerlof1999} and GRB~210619B \autocite{Oganesyan2023}; and, approximately, a factor of 15 lower than the parameter value taken in the calculations by \textcite{Draine2002}. 

\begin{figure*}[h]
\centering
\includegraphics[width = 0.8\textwidth]{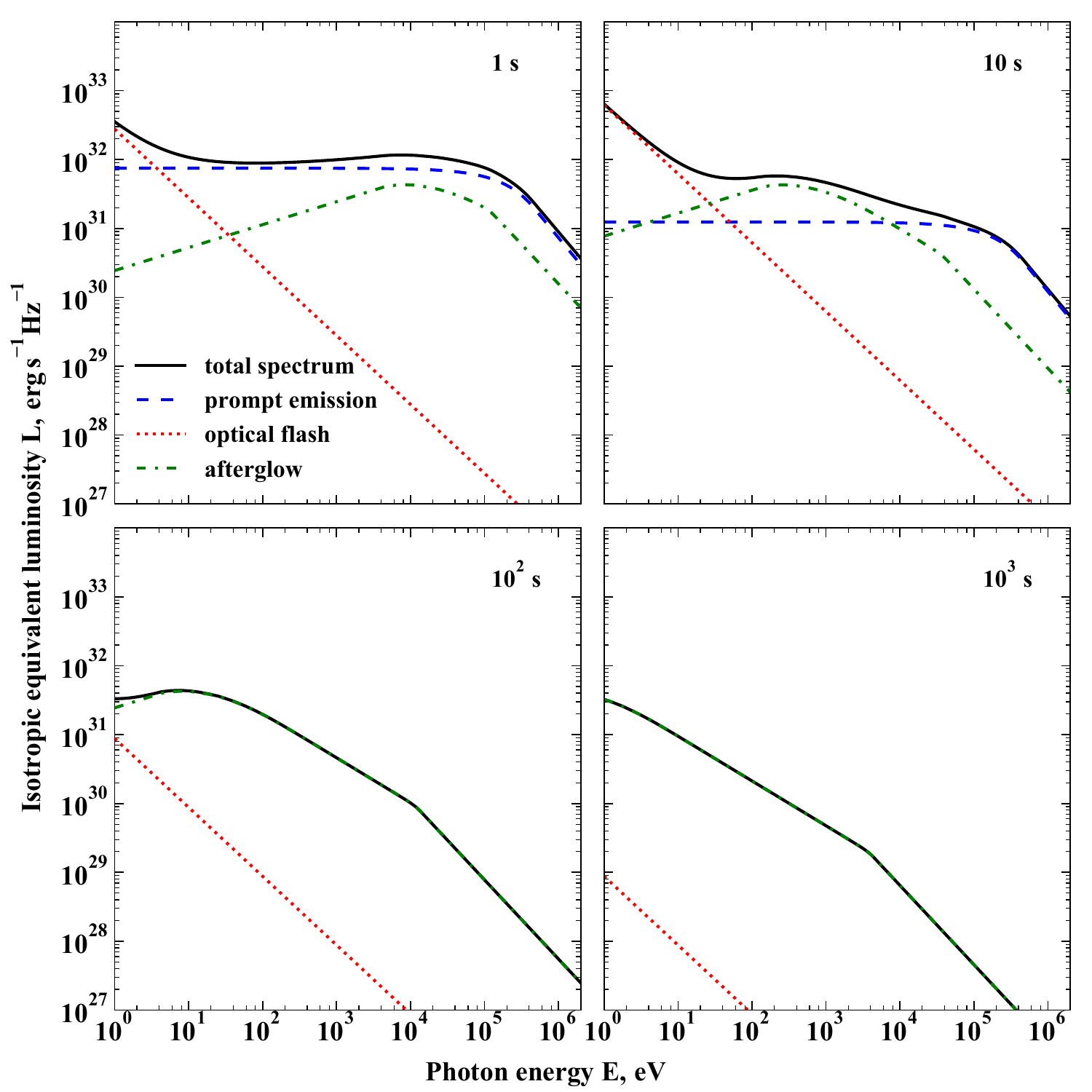}
\caption{\rm Isotropic equivalent luminosity for different components of gamma-ray burst emission (the prompt emission, the optical flash, and the afterglow). The photon energy in eV is along the horizontal axis, the isotropic equivalent luminosity (e.g., under the assumption that the radiation is emitted into the total solid angle of $4\pi$) is along the vertical axis in erg~s$^{-1}$~Hz$^{-1}$. The graphs show the isotropic equivalent luminosity at four different times: 1, 10, $10^2$, $10^3$~s.}
\label{fig1}
\end{figure*}

Figure~\ref{fig1} shows the isotropic equivalent luminosity of the gamma-ray burst emission components (the emission of the prompt phase, the optical flash, and the afterglow) at times 1, 10, $10^2$, $10^3$~s after the start of the gamma-ray burst. The optical flash is the most intense emission component in the optical and UV wavelength ranges at times $t < 20$~s.
The luminosities of the reverse and forward shocks depend on microphysical parameters of the jet. In particular, the luminosity $L_{0}$ is proportional to the fraction of the energy that is in the magnetic field $\varepsilon_{\rm B}$ \autocite{Nakar2005}. While estimates of the parameter $\varepsilon_{\rm B}$ for the forward shock vary over a wide range for various gamma-ray bursts, $10^{-8} \lesssim \varepsilon_{\rm B} \lesssim 10^{-1}$ \autocite{Santana2014}.

\subsection{The gas--dust cloud model}
In our numerical model, the gamma-ray burst radiation propagates in a homogeneous cloud located at some small distance from the burst progenitor star. The source of the gamma-ray burst radiation is located at the coordinate origin and is assumed to be a point-like in the calculations -- the distance travelled by the jet in the fireball model during the time period in question ($10^5$~s) is much smaller than the ionization radius of the cloud \autocite{Lazzati2001}. 
The gas density in units of the total number density of hydrogen nuclei $n_{\rm H,tot}$ is taken equal to $10^3$~cm$^{-3}$. At the initial moment of time, hydrogen atoms are bound in H$_2$ molecules. The parameters of the molecular cloud are provided in Table~2. The chemical elements H, He, C, N, O, Ne, Mg, Si, S, Fe are taken into account in the simulations. The abundance of He is taken to be 0.09. The abundances of chemical elements correspond to the solar abundances, and the correction is made for the metallicity, $[M/H] \equiv {\rm log_{10}} (Z/Z_{\odot})$.
The results presented in figures correspond to the metallicity $[M/H] = 0$; the results of the calculations for metallicities $[M/H] = -1$ and $-2$ are discussed in the text of the paper. The abundances of chemical elements for $[M/H] = 0$ are given in Table~3 \autocite{Lodders2021}. At the initial time moment atoms and H$_2$ molecules are in the neutral state, the number density of H atoms is equal to 0. 

At a distance of 1~pc or less, the typical time of the gas ionization by gamma-ray burst radiation is $1-3$~s. We consider distances $R \geq R_{\rm min} = 1$~pc from the burst location to reduce the computation time of the gas ionization. The cloud is divided into spherical shells of equal thickness $\Delta R$. Each shell is distinguished by the inner radius $R_j$ (the distance from the burst source to the inner boundary of the shell), and mean radius $\overline{R}_j = (R_j + R_{j + 1})/2$.
The shell thickness is taken equal to $\Delta R = 5 \times 10^{16}$~cm. For each radius $R$ the retarded time is determined $t_{r} = t - R/c$, where $t$ is the time passed since the gamma-ray burst onset. The retarded time $t_{r} = 0$ corresponds to the time moment when the gamma-ray burst radiation front reaches the cloud shell at distance $R$.

\begin{table*} 
	\caption{\normalsize Parameters of the gas--dust cloud.}
	\centering 
	\begin{tabular}{L{0.41\linewidth} L{0.2\linewidth}} 
	\toprule
	Distance from the gamma-ray burst progenitor & 1~pc \\
	Gas density in the cloud, $n_{\rm H,tot}$ & $10^3$~cm$^{-3}$ \\
	Initial ortho-to-para H$_2$ ratio & 1 \\
	Micro-turbulence speed, $v_{\rm turb}$ & 5~km/s \\
	Metallicity, $[M/H]$ & 0, $-1$, $-2$ \\
	Fraction of metals (Mg, Si, Fe) in dust & 0.99 \\
	Dust--gas mass ratio (for $[M/H] = 0$) & 0.004 \\
    \bottomrule
	\end{tabular}
\end{table*}

\begin{table} 
	\caption{\normalsize Abundances of chemical elements relative to the hydrogen nuclei, $[M/H] = 0$.}
	\centering
	\begin{tabular}{l l}
	\toprule
	He & 0.09 \\
	C  & $2.95 \times 10^{-4}$ \\
	N  & $7.08 \times 10^{-5}$ \\
	O  & $5.37 \times 10^{-4}$ \\
	Ne & $1.41 \times 10^{-4}$ \\
	Mg & $3.16 \times 10^{-5}$ \\
	Si & $3.16 \times 10^{-5}$ \\
	S  & $1.41 \times 10^{-5}$ \\
	Fe & $3.16 \times 10^{-5}$ \\
	\bottomrule
	\end{tabular}
	\label{tab:distcounts}
\end{table}

\subsection{Calculation of optical depths and photoionization and photodissociation rates}
In the calculations we take into account the range of electromagnetic radiation from 1~eV to 1~MeV. This range is divided into intervals whose lengths follow a geometric progression. The total number of intervals is 600, 100 intervals per each order of magnitude of frequency. The optical depth of the cloud shell $j$ is the sum of optical depths due to several processes:

\begin{equation}
\begin{gathered}
\Delta\tau_{j} = \Delta\tau_{{\rm dust},j} + \sum_{\alpha} \Delta\tau_{{\rm H_2},\alpha,j} + \sum_{\rm X} \sum_{a} \Delta\tau_{{\rm X^{+a}},j} + \\
+ \Delta\tau_{{\rm H_2},j} + \Delta\tau_{{\rm H^+_2},j} + \Delta\tau_{{\rm H},j} + \Delta\tau_{{\rm He},j} +\\
+ \Delta\tau_{{\rm He^+},j} + \Delta\tau_{{\rm e^-},j},
\end{gathered}
\end{equation} 

\noindent
where $\Delta\tau_{{\rm dust},j}$ is the optical depth of the cloud shell arising from the absorption by dust, $\Delta\tau_{\alpha,j}$ is the optical depth of the shell due to bound--bound transitions of the Lyman and Werner bands of H$_2$ molecule, $\Delta\tau_{{\rm X^{+a}},j}$ is the optical depth resulted from the photoionization and the Compton ionization of metal ion $X$ with charge $+a$ in the gas phase (where $X$ is one of C, N, O, Ne, Mg, Si, S, Fe), $\Delta\tau_{{\rm e^-},j}$ is the Compton scattering by free electrons.
The optical depths $\Delta\tau_{{\rm H_2},j}$, $\Delta\tau_{{\rm H^+_2},j}$, $\Delta\tau_{{\rm H},j}$, $\Delta\tau_{{\rm He},j}$ and $\Delta\tau_{{\rm He^+},j}$ take into account the photoionization and the Compton ionization processes for H, H$_2$, H$^{+}_{2}$, He, He$^{+}$, and the photodissociation processes for H$_2$ and H$^+_2$. The optical depth from the gamma-ray burst progenitor to the inner boundary of the cloud shell $j$ is equal to the sum of contributions to the optical depth of all cloud shells preceding the shell $j$:

\begin{equation}
\tau_{j-1}(\nu,t_r) = \sum_{i=1}^{j-1} \Delta \tau_i(\nu,t_r),
\end{equation}

\noindent
where optical depths are calculated at the same retarded time $t_{r}$ when summing over the shells.

The optical depth of the cloud shell $j$ for each of the processes of photoionization, photodissociation, and the Compton ionization equals:

\begin{equation}
\Delta\tau_{X,j} = n_{X}^{j} \sigma_{X}(\nu)\Delta R,
\end{equation}

\noindent
where $n_{X}^{j}$ is the number density of target ions (molecules) of type $X$ in the cloud shell $j$, $\sigma_X(\nu)$ is the cross section of the reaction in question. The rate of ionization (or photodissociation) of ions (molecules) of type $X$, i.e. the number of reaction events per unit time and per specimen of type $X$, averaged over the cloud shell is calculated using equation \autocite{Draine2002}:

\begin{equation}
k_{X}^{j} = \frac{1}{4\pi \overline{R}^2_j}\int\limits^{\infty}_{\nu_0} d\nu \frac{L(\nu)}{h\nu} e^{-\tau_{j-1}}\frac{1-e^{-\Delta\tau_j}}{\Delta \tau_j} \sigma_X(\nu).
\label{eq_ion_rate_coeff}
\end{equation}

\noindent
Let $n_{X,a}^{j}$ be the number density of ions of chemical element $X$ with charge $+a$ in the cloud shell $j$. The time-evolution of the number density of ions $X^{+a}$ is determined by the system of equations:

\begin{equation}
\begin{gathered}
\frac{d n_{X,a}^{j}}{dt} = -k_{X,a}^{j} n_{X,a}^{j} + \\
+ \sum_{b = 0}^{a-1} P_{X,a,b} k_{X,b}^{j} n_{X,b}^{j} + k^{j}_{{\rm vap}, X, a},
\end{gathered}
\end{equation}

\noindent
where $k_{X,a}^{j}$ is the photoionization rate (including the Compton ionization) of ions $X^{+a}$, $P_{X,a,b}$ is the probability of the emission of $a-b-1$ electrons as a result of the Auger effect after the photoionization of the ion $X^{+b}$, $k^{j}_{{\rm vap}, X, a}$ is the release rate of ions arising from the thermal sublimation of dust grains (see the next section).
The photoionization rate of ions $X^{+b}$ is calculated for each of the inner electron shells and, based on these data, the probabilities $P_{X,a,b}$ are calculated. The ionization of ions in the collisions with electrons may be neglected as we investigate cloud regions behind the ionization front where the ionization fraction is relatively low.
In particular, the cross section of collisional ionization of CI is $\sim 10^{-16}$~cm$^2$ at electron energies $E \sim 15-10^3$~eV \autocite{Brook1978}. The typical ionization time is $\sim 10^7$~s for electron number density 1~cm$^{-3}$ and energy $E = 1$~keV. This time is much longer than the time interval considered in our work. The ion recombination rates are also low for the gas densities in question \autocite{Perna1998,Lazzati2001}.

\subsection{The dust model}
At the initial time moment dust grains have the same radius $r_{\rm d} = 0.1$~$\mu m$. The density of the dust material is $\rho_{\rm d} = 3.5$~g/cm$^{3}$ and its chemical composition is taken to be MgFeSiO$_4$. The dust-to-gas mass ratio is determined by the metallicity and dust depletion of metals. The dust depletion of metals is assumed to be 0.99 (an arbitrary value close to 1). For the metallicity $[M/H] = 0$, the dust-to-gas mass ratio equals to 0.004 at the initial time moment (we take into consideration only silicate dust, and metal ions that are not locked in dust grains are in the gas phase). 

The optical depth of the cloud shell due to the absorption and scattering of radiation by dust grains of radius $r_{\rm d}$ is:

\begin{equation}
\Delta \tau_{{\rm dust},j} = \pi r_{\rm d}^2 \left[Q_{\rm abs}(r_{\rm d}, \nu) + Q_{\rm sca}(r_{\rm d}, \nu) \right] n_{\rm d} \Delta R,
\label{eq_tau_dust}
\end{equation}

\noindent
where $Q_{\rm abs}(r_{\rm d}, \nu)$ and $Q_{\rm sca}(r_{\rm d}, \nu)$ are the efficiency factors for absorption and scattering, $n_{\rm d}$ is the number density of dust grains in the gas--dust cloud. The efficiency factors for absorption and scattering of light by dust grains for optical and UV wavelength ranges were calculated by \textcite{Laor1993,Weingartner2001}.
The absorption and scattering efficiency factors in the X-ray range were calculated by \textcite{Draine2003}. The data on these factors for photon energies $h\nu \leq 1240$~eV were available at the web page of Prof. Draine\footnote{\url{https://www.astro.princeton.edu/~draine/dust/dust.html}}. At higher photon energies, the extrapolation of the efficiency factors by power-law function is used \autocite{Draine2002,Draine2003}.
The ionization of dust grains in the process of radiation absorption is not taken into account. The photon wavelength in the X-ray range is small compared to the dust grain sizes, and the scattering occurs predominantly forward. For scattering angles less than $\theta_{\rm min}$, it can be considered that the photon does not leave the main gamma-ray burst radiation pulse. According to the estimates by \textcite{Draine2002}, $\theta_{\rm min} \sim 5-10'$. The term corresponding to the scattering is omitted in the equation (\ref{eq_tau_dust}) at photon energies $h \nu > 5$~keV \autocite{Draine2002,Draine2003}.

We use an approximate formula for the thermal sublimation rate of dust grains \autocite{Guhathakurta1989,Waxman2000}: 

\begin{equation}
\frac{dr_{\rm d}}{dt} = -\left(\frac{m_{\rm d}}{\rho_{\rm d}} \right)^{1/3} \nu_0 \exp\left(-\frac{B}{k_{\rm B} T_{\rm d}}\right),
\end{equation}

\noindent
where $T_{\rm d}$ is the dust grain temperature, $\rho_{\rm d}$ is the density of dust grain material, $m_{\rm d}$ is the mean atomic mass of the dust material, $B/k_{\rm B} = 7 \times 10^4$~K is the chemical binding energy per atom, and $\nu_0 = 10^{15}$~s$^{-1}$. The dust temperature $T_{\rm d}$ is determined by the equation for the balance of grain heating and cooling rates:

\begin{equation}
C_{\rm d} \frac{dT_{\rm d}}{dt} = G_{\rm GRB} + G_{\rm rad} + G_{\rm vap}, 
\end{equation}

\noindent
where $G_{\rm GRB}$ is the heating rate of dust grains by the radiation of the gamma-ray burst, $G_{\rm rad}$ is the cooling rate of dust grains through their thermal emission, $G_{\rm vap}$ is the cooling rate of dust grains as a result of their sublimation, $C_{\rm d}$ is the heat capacity of dust grains. The heating rate of a dust grain by the radiation of the gamma-ray burst in the middle of shell $j$ is:

\begin{equation}
G_{\rm GRB} = \frac{1}{4\pi\overline{R}^2_{j}} \int\limits_{0}^{\infty} d\nu L(\nu) e^{-\tau_{j-1} -0.5\Delta\tau_j} Q_{\rm abs} \pi r^2_{\rm d}.
\label{eq_dust_rad_heating}
\end{equation}

\noindent
The cooling rate of the dust grain due to thermal emission is:

\begin{equation}
G_{\rm rad} = -4\pi \int\limits_0^{\infty} B(\nu, T_{\rm d}) Q_{\rm abs} \pi r^2_{\rm d},
\end{equation}

\noindent
where $B(\nu, T_{\rm d})$ is the blackbody radiation intensity. The cooling rate of the dust grain resulted from its sublimation is:

\begin{equation}
G_{\rm vap} = -4\pi r^2_{\rm d} \left\vert \frac{d r_{\rm d}}{dt} \right\vert \frac{B \rho_{\rm d}}{m_{\rm d}}.
\end{equation}

\noindent
The heat capacity of the dust grain $C_{\rm d}$ is approximated by the formula $3k_{\rm B}N$, where $N$ is the number of atoms in one dust grain. For silicate dust grains, the error of this approximation is less than 10 per cent at dust temperatures $T_{\rm d} > 600$~K \autocite{Draine2001}. We do not take into account the destruction of dust grains due to ion field emission. According to the estimates by \textcite{Draine2002,Perna2002}, this mechanism of grain destruction takes place at distances $R\lesssim 1-5$~pc from the burst progenitor (the distance estimate depends on the X-ray photon fluence).
Meanwhile, the heating of dust grains by absorption of radiation in optical and UV wavelength ranges brings to the thermal sublimation of grains at distances $R \lesssim 10$~pc (for the parameters of gamma-ray burst and gas--dust cloud in question). 

The release rate of atoms into the gas phase as a result of dust grain sublimation is:

\begin{equation}
k_{{\rm vap},X,a} = 4 \pi r^2_{\rm d} \left\vert \frac{d r_{\rm d}}{dt} \right\vert \frac{\rho_{\rm d} n_{\rm d}}{m_{\rm d}} \frac{l}{7},
\end{equation} 

\noindent
where $l = 1$ for Mg, Fe, and Si; $l = 4$ for O; $l = 0$ for other chemical elements and ions with charge $a > 0$ (it is assumed that neutral atoms evaporate off of the grain surface).

\subsection{The bound--bound H$_2$ transitions in the Lyman and Werner bands}
The system of equations for the populations of ro-vibrational levels of the ground electronic state of H$_2$ molecule is:

\begin{equation}
\begin{gathered}
\frac{dn_l}{dt} = \sum_{m \neq l} \beta_{lm} n_m - n_{l}\left( \sum_{m \neq l} \beta_{ml} + \beta_{{\rm diss}, l} \right) -\\
- n_l k_{{\rm H_2},l} + \sum_{m > l} A_{lm} n_{m} - n_{l} \sum_{m < l} A_{ml},
\end{gathered}
\label{eq_h2_stat_equil}
\end{equation}

\noindent
where $n_l$ is the number density of H$_2$ molecules at an energy level $l$, $\beta_{lm}$ is the effective rate of the transition $m \to l$ due to the photoexcitation of electronic states of H$_2$ and the radiative transition back to the ground electronic state, $\beta_{{\rm diss}, l}$ is the H$_2$ dissociation rate as a result of radiative decay of excited states into ro-vibrational continuum of the ground electronic state,
$k_{{\rm H_2}, l}$ is the H$_2$ destruction rate in the processes of photoionization, direct photodissociation, the Compton ionization, and $A_{lm}$ are the Einstein coefficients for the spontaneous radiative transition from the energy level $m$ to $l$ (within the ground electronic state of the molecule). The dependence of the parameter $k_{{\rm H_2}}$ on the H$_2$ energy level is taken into account only for the direct photodissociation \autocite{Gay2012}. The transitions between energy levels due to collisions of molecules may be neglected for gas densities and time period being considered in our model \autocite{Bossion2018,Wan2018}. 

The effective rate of the transition from the level $m$ to $l$ due to the photoexcitation of electronic states of H$_2$ is:

\begin{equation}
\beta_{lm} = \sum_{u} \frac{A_{lu}}{A_{{\rm tot},u}} \zeta_{um},
\end{equation}

\noindent
where $\zeta_{um}$ is the photoexcitation rate from the level $m$ to the level $u$ of the excited electronic state, $A_{lu}$ is the probability of the spontaneous radiative transition from the level $u$ to $l$, $A_{{\rm tot},u}$ is the total probability of the level decay as a result of the spontaneous emission that also includes the probability of the transition to the ro-vibrational continuum of the ground electronic state $A_{{\rm vc}, u}$,

\begin{equation}
A_{{\rm tot}, u} = \sum_{l} A_{lu} + A_{{\rm vc}, u}.
\end{equation}

\noindent
The H$_2$ photodissociation rate in the Solomon process is:

\begin{equation}
\displaystyle
\beta_{{\rm diss},m} = \sum_{u} \frac{A_{{\rm vc}, u}}{A_{{\rm tot}, u}} \zeta_{um}.
\end{equation}

\noindent
The frequency width of each line from the Lyman and Werner bands is much narrower than the width of the intervals into which the photon frequency scale is divided in the calculations. The attenuation of radiation intensity is taken into account using the equivalent width method. Let the radiative transition $\alpha$ couple the energy level $l$ of the ground electronic state and the energy level $u$ of the excited electronic state of H$_2$ molecule. The dimensionless
equivalent width of the line $\alpha$ at distance $R_{j}$ from the burst source is \autocite{Draine2011}:

\begin{equation}
\displaystyle
W_{ul}(R_j) = \int \frac{d \nu}{\nu} \left\{ 1 - \exp\left[-N_{l}(R_j)\sigma_{ul}(\nu) \right] \right\},
\label{eq_h2_eq_width}
\end{equation} 

\noindent
where $\sigma_{ul}(\nu)$ is the cross section of photon absorption in the transition $\alpha$ as a function of photon frequency, $N_{l}(R_{j})$ is the column density of H$_2$ molecules at the energy level $l$:

\begin{equation}
N_{l}(R_{j}) =\sum_{i=1}^{j-1} n_{l}^{i} \Delta R,
\end{equation} 

\noindent
where $n_{l}^{i}$ is the number density of H$_2$ molecules at the energy level $l$ in the cloud shell $i$. The values of $n_{l}^{i}$ in cloud shells are considered at the same retarded time $t_{r}$. The photoexcitation rate of H$_2$ molecules as a result of photon absorption in the transition $\alpha$ in the cloud shell $j$ is equal to \autocite{Draine1996,Draine2002}:

\begin{equation}
\zeta_{ul}(N_{l}) = \frac{L(\nu) e^{-\tau_{j-1}}}{4 \pi \overline{R}^2_j h} \frac{\Delta W_{ul}}{\Delta N_l},
\label{eq_h2_photopump_rate}
\end{equation}

\noindent
where the increments in column density $\Delta N_l$ and equivalent width $\Delta W_{ul}$ contributed by shell $j$ are:

\begin{equation}
\Delta N_l = n_{l}^{j} \Delta R, \quad \Delta W_{ul} = W_{ul}(R_{j+1}) - W_{ul}(R_{j}).
\end{equation}

\noindent
The contribution of the transition $\alpha$ to the optical depth averaged over the frequency interval $[\nu_{0}, \nu_{0} + \Delta\nu]$ to which the transition $\alpha$ belongs to is \autocite{Draine1996,Draine2002}:

\begin{equation}
\Delta \tau_{\alpha,j} = \frac{\Delta W_{ul}}{{\rm ln} (1 + \Delta\nu/\nu_{0})}.
\end{equation}

\noindent
In the calculations of equivalent widths, the numerical approximations given by \textcite{Rodgers1974} are used.

The calculations of energy level populations are performed for 301 ro-vibrational levels of the ground electronic state $X^1\Sigma^+_g$ of H$_2$ molecule. The energies of ro-vibrational levels of the ground electronic state $X^1\Sigma^+_g$ and the excited electronic states $B^1\Sigma^+_u$ and $C^1\Pi^{\pm}_u$, as well as the Einstein coefficients for transitions $X^1\Sigma^+_g \leftarrow B^1\Sigma^+_u$ and $X^1\Sigma^+_g \leftarrow C^1\Pi^{\pm}_u$ are taken from the files with the data of the CLOUDY code \autocite{Shaw2005,Ferland2017}.
The Einstein coefficients for transitions within the ground electronic state of the molecule are taken from \textcite{Wolniewicz1998}. The absorption of radiation in transitions of atoms and molecules other than H$_2$ is not taken into account.  

\subsection{Numerical calculations}
At photon energies below $13.6$~eV (the ionization threshold of H atom), the main contribution to the optical depth comes from the dust absorption, the photoionization of metal atoms with a low ionization potential (CI, MgI, SiI, SI, FeI), and the absorption of H$_2$ in transitions of the Lyman and Werner bands. At photon energies above $13.6$~eV, the photoionization of H, H$_2$ and He, the direct photodissociation of H$_2$ make a major contribution to the optical depth.
At photon energies $0.5 \lesssim h\nu \lesssim 10$~keV, the absorption and scattering by dust and the photoionization of metal ions in the gas phase become significant again. At these photon energies, the photoionization of inner electron shells of metal ions makes a major contribution to the photoionization cross sections of ions. The Compton ionization of H, H$_2$ and He is the main photon absorption process at higher energies.
The contribution of the various constituents of the interstellar medium to the optical depth changes during the photoionization and photodissociation of H$_2$ and the vaporization of dust grains. In particular, the atomic hydrogen is produced as the result of H$_2$ destruction, and the photoionization of H atoms mainly contributes to the optical depth at photon energies $13.6 \leq E \leq 15.4$~eV.

The scheme of our numerical simulations is the same as in the papers by \textcite{Perna2000,Draine2002}. The calculations of the time dependence of the number densities of ions and chemical compounds are carried out sequentially for each cloud shell. We solve the system of differential equations for the energy level populations of H$_2$ molecule, the number densities of ions and chemical compounds, and the radius and temperature of dust grains.
The system of differential equations is solved using the numerical code SUNDIALS CVODE v5.7.0 \autocite{Hindmarsh2005,Gardner2022}. The maximal model time for which the simulations are carried out is $t_{\rm max} = 10^5$~s. The integration of the system of equations is stopped before this time if full gas ionization is reached. The criterion for the full gas ionization is chosen to be the following:

\begin{equation}
\begin{gathered}
n_{\rm H_2} + n_{\rm H^+_2} + n_{\rm H} +n_{\rm He} + n_{\rm He^+} +\\ 
+\sum_{\rm X} \sum_{a < a_{\rm max}} n_{\rm X^{+a}} < 10^{-8} n_{\rm H,tot},
\end{gathered}
\end{equation}

\noindent
where metal ions with a charge below the maximal charge for the particular ion are taken into account in the sum over ion charges. In solving the system of differential equations for the cloud shell $j$, we use optical depths $\tau_{j-1}(\nu, t_r)$ and column densities $N_l(R_{j}, t_r)$ of level populations of H$_2$ inferred in the calculations for cloud shells $i < j$.

\begin{figure*}[h]
\centering
\includegraphics[width = 0.6\textwidth]{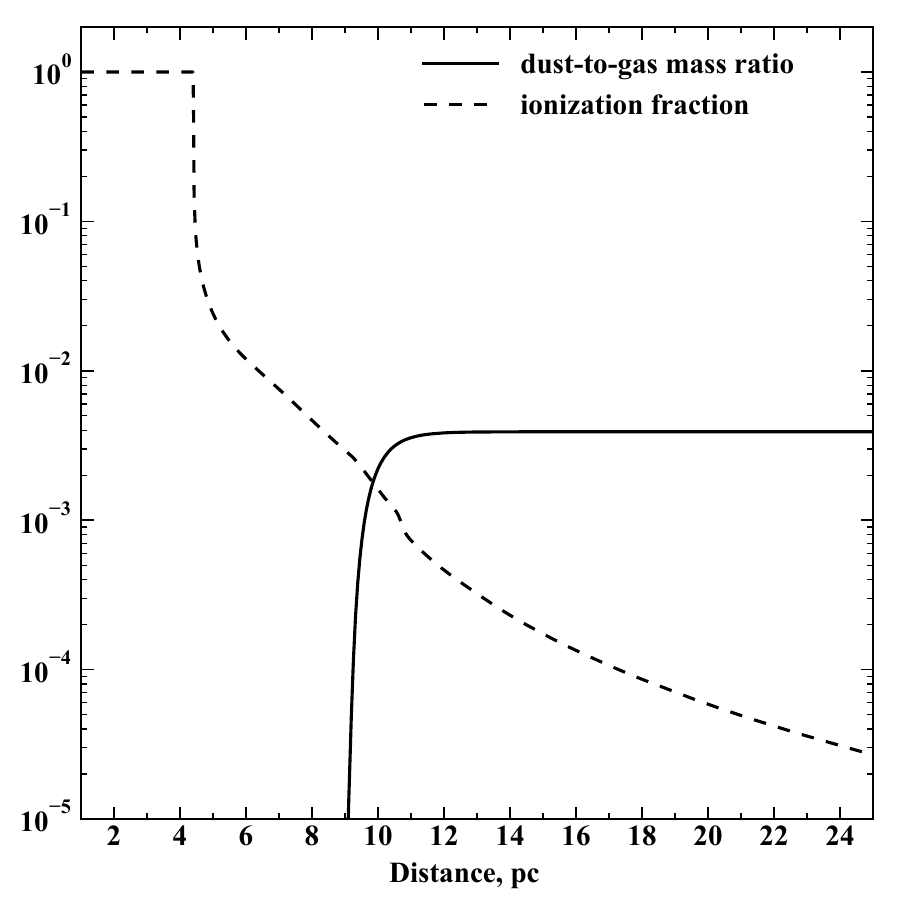}
\caption{\rm The gas ionization fraction $x_{\rm i}$ and dust-to-gas mass ratio as a function of the distance from the gamma-ray burst progenitor. The values of the parameters are given for each distance at the same retarded time $t_{\rm max} = 10^5$~s.}
\label{fig2}
\end{figure*}

\section{Results}
\subsection{The ionization structure of the cloud}
Let us define the gas ionization fraction:

\begin{equation}
x_{\rm i} = \frac{n_{\rm e}}{n_{\rm e, tot}},
\end{equation}

\noindent
where $n_{\rm e}$ is the number density of electrons in the gas, and $n_{\rm e, tot}$ is the total number of electrons per unit volume (both free electrons and electrons bound in ions and molecules). Figure~\ref{fig2} shows the gas ionization fraction $x_{\rm i}$ and dust-to-gas mass ratio as a function of the distance from the gamma-ray burst progenitor at the time $t_{\rm max} = 10^{5}$~s after the burst onset (at the same retarded time).
There is a boundary between the fully ionized gas and neutral gas at a distance $R_{\rm ion} = 4.4$~pc from the gamma-ray burst progenitor. The gas ionization fraction at this boundary drops by a factor of $\approx 30$ and further out decreases slowly with the distance. In the region where $x_{\rm i} << 1$, the main source of electrons in the gas is the ionization of metal atoms.

The molecular gas is opaque to radiation at photon energies above $13.6$~eV. Therefore, the dust is sublimated as a result of its heating by radiation in the optical and UV wavelength ranges (see also \textcite{Draine2002}). The sublimation rate of dust grains has a strong dependence on the dust temperature, which, in turn, depends on the radiation intensity.  
The radiation intensity is not high enough at distances $R > R_{\rm d} \approx 9-10$~pc, and dust grains do not sublimate. There is complete sublimation of dust grains at smaller distances, with the characteristic sublimation time being about $3-5$~s from the start of the gamma-ray burst (dust grains are heated and sublimated by UV radiation of the optical flash of the gamma-ray burst).
Thus, the ionizing radiation at distances $R < R_{\rm d}$ is not shielded by the dust that has already been sublimated within the first seconds of the gamma-ray burst. In this case, the ionization fraction of the gas depends on the fluence of the ionizing radiation, but not on the functional dependence of the radiation flux on time \autocite{Lazzati2002,Lazzati2003}.
The drop in the ionization fraction at a distance of $\approx 9-11$~pc is related to the decrease in the number density of metals in the gas phase (O, Mg, Si, Fe) at $R > R_{\rm d}$, and to the transition of carbon ions to the neutral state (CII/CI transition). The contribution of the Compton ionization to the number density of electrons in the gas increases with the distance from the gamma-ray burst progenitor, and is about 15 per cent at a distance of 25~pc. 

\begin{figure*}[h]
\centering
\includegraphics[width = 0.6\textwidth]{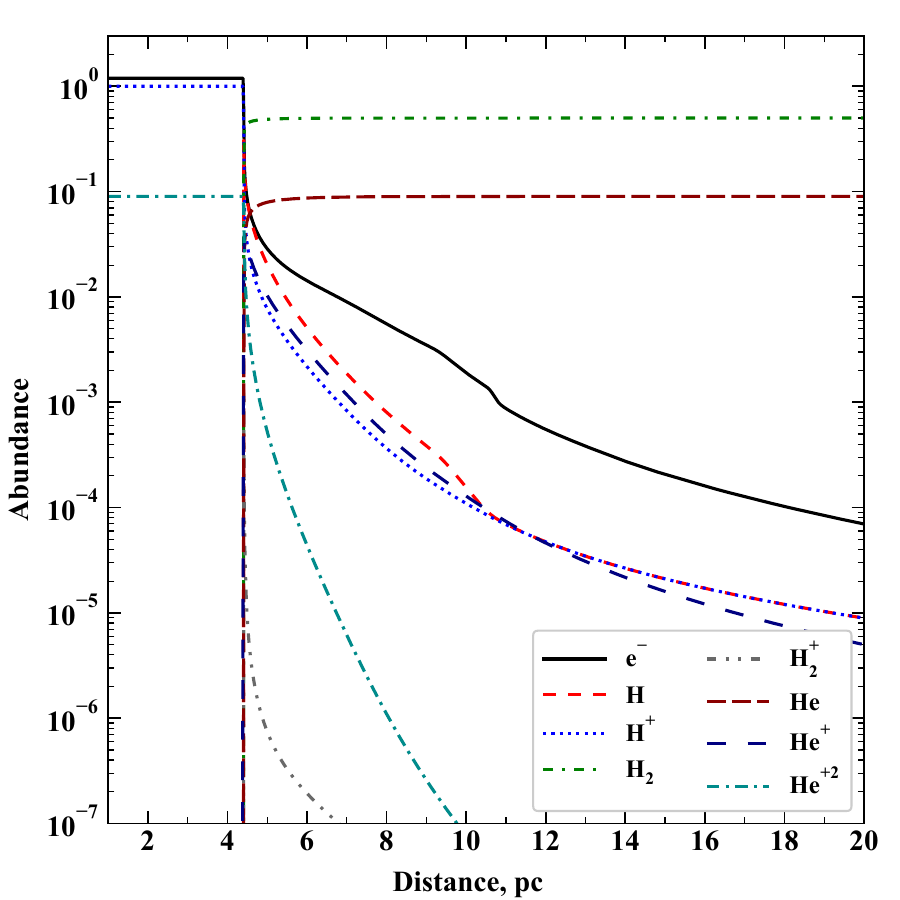}
\caption{\rm The abundances of H, He, H$_2$ and their ions as a function of the distance to the gamma-ray burst progenitor (at the retarded time $t_{\rm max} = 10^5$~s).}
\label{fig3}
\end{figure*}

The abundances of ions and molecules are calculated using the formula:

\begin{equation}
x_{\rm A} = \frac{n_{\rm A}}{n_{\rm H,tot}},
\end{equation}

\noindent
where $n_{\rm A}$ is the number density of atoms (molecules) of type $A$. Figure~\ref{fig3} presents the abundances of H$_2$ and H$_2^+$ molecules, ions H, H$^+$, He, He$^{+}$, He$^{+2}$ and electrons as a function of the distance to the gamma-ray burst progenitor. The ratio of number densities $n_{\rm H}/ n_{\rm H_2}$ is equal to $\sim 1$ in a narrow region with a size $\sim 0.01$~pc at the boundary ionized/neutral gas in the cloud.
The destruction reactions of molecular hydrogen are (i) the photoionization of H$_2$; (ii) the dissociative photoionization of H$_2$; (iii) the Compton ionization; (iv) the direct photoionization of H$_2$; (v) the Solomon process. The direct photodissociation and the Solomon process of H$_2$ destruction are efficient at distances smaller than the dust sublimation radius (they contribute $40-50$ per cent to the total amount of destroyed H$_2$ at $R_{\rm ion} < R < R_{\rm d}$). 
At larger distances, the dissociation rate of H$_2$ drops rapidly with the distance (the UV radiation is shielded by the dust), and the destruction of H$_2$ is mainly due to ionization. The main destruction mechanism of H$_2^+$ is the photodissociation with the formation of H and H$^+$. Therefore, the number densities of H and H$^+$ are equal at $R > R_{\rm d}$ (Fig.~\ref{fig3}). At distances $R > 13$~pc, the Compton ionization becomes the main H$_2$ destruction process.

The gas metallicity does not affect the radius of the gas ionization front. The dust-destruction radius increases with decreasing metallicity: $R_{\rm d} \approx 12-13$~pc at $[M/H] = -1$, $-2$.

\begin{figure*}[h]
\centering
\includegraphics[width = 0.9\textwidth]{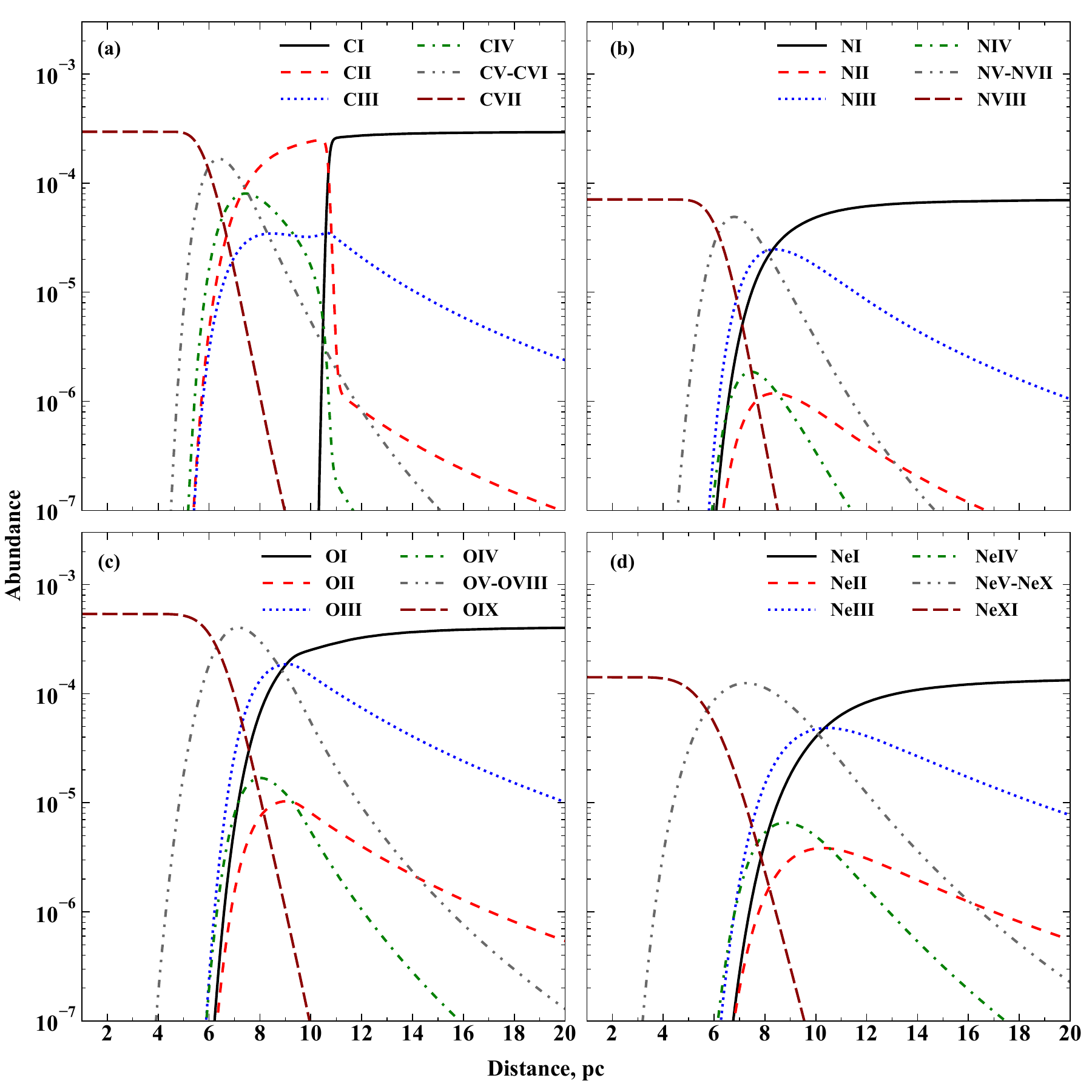}
\caption{\rm The abundances of C, N, O and Ne ions in various states of ionization as a function of distance (at the retarded time $t_{\rm max} = 10^5$~s). The notation NV--NVII means the total abundance of ions from NV to NVII, and similarly for other ions.}
\label{fig4}
\end{figure*}

\subsection{The abundances of metal ions}
Figure~\ref{fig4} shows the abundances of metal ions C, N, O, and Ne as a function of distance. The ionization potential of the outer electron shell of the neutral carbon atom CI is 11.2~eV. The UV photons with energies lower than the ionization potential of atomic hydrogen, $E < 13.6$~eV, make a major contribution to the ionization of CI atoms. At a certain distance from the burst progenitor the fluence of the UV radiation is not high enough to fully ionize CI atoms, and carbon remains predominantly neutral.
The transition CII/CI is abrupt and takes place at a distance $R_{\rm C} \approx 10-11$~pc, behind the dust-destruction front (Fig.~\ref{fig4}). At distances $R > R_{\rm C}$, the CI atoms are photoionized through the ionization of electron K-shell of an atom. For neutral NI, OI, NeI atoms, the ionization potential of the outer electron shell is approximately equal to or higher than 13.6~eV. The ionization of the electron K-shells makes a major contribution to the photoinization of these atoms at all distances from the gamma-ray burst source.
One Auger electron is ejected with a probability close to 1 as a result of ionization of electron K-shells of CI, NI, OI, NeI \autocite{Kaastra1993}. As a consequence, neutral NI atoms turn directly into NIII, and the same happens to CI, NI, NeI atoms (for CI atoms this is true at distances $R > R_{\rm C}$). The photoionization of CII (CIII), NIII, OIII and NeIII also occurs through the ionization of the K-shells of ions which is accompanied by the emission of one Auger electron.
Thus, the number density of ions in the state of ionization II is much lower than the number density of ions in the ionization state III, and the number density of ions in the ionization state IV is much lower than the number density of ions in higher ionization states (for carbon this is true at distances $R > R_{\rm C}$), see Fig.~\ref{fig4}.

\begin{figure*}[h]
\centering
\includegraphics[width = 0.9\textwidth]{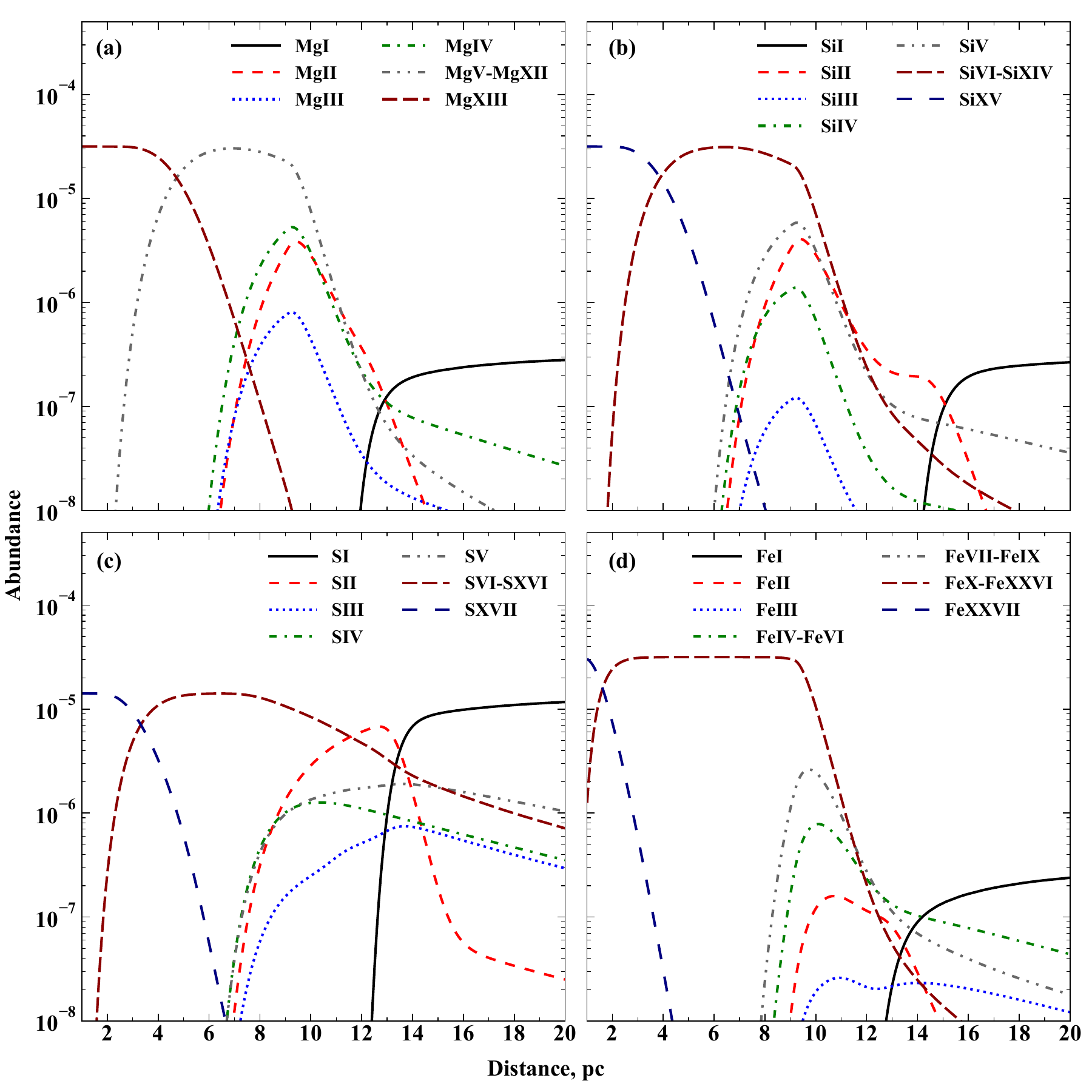}
\caption{\rm Same as in Fig.~\ref{fig4}, only for Mg, Si, S and Fe ions.}
\label{fig5}
\end{figure*}

Figures~\ref{fig5}a, \ref{fig5}b and \ref{fig5}d show the abundances of Mg, Si and Fe ions as a function of distance. In the region $R \lesssim R_{\rm d}$, Mg, Si, Fe atoms release into the gas phase due to the sublimation of dust grains. The ionization potentials of outer electron shells of MgI, SiI and FeI atoms are 7.65, 8.15 and 7.9~eV, respectively.
These atoms are photoionized through the absorption of UV photons. The ionization potentials of MgII, SiII and FeII are 15, 16.4 and 16.2~eV, respectively. These ions are ionized via the absorption of photons in the X-ray range and the photoionization of their inner electron shells. In this case, the photoionization is accompanied (with a high probability) by the emission of 1 or 2 Auger electrons for MgII, $1-3$ Auger electrons for SiII, and $1-6$ Auger electrons for FeII.
The Mg, Si and Fe atoms are predominantly in the neutral state at distances $R \geq 13-15$~pc from the gamma-ray burst progenitor (Fig.~\ref{fig5}). At these distances, the most probable ionization route of MgI is the photoionization of the electron K-shell, which is accompanied by the emission of two Auger electrons. Therefore, at distances $R > R_{\rm Mg} \approx 13$~pc, the next most abundant ion after MgI is MgIV, and the number density of Mg ions in the remaining states of ionization is lower then the number density of MgIV (Fig.~\ref{fig5}).
Similarly, SiV is the second most abundant ion after SiI at distances $R > R_{\rm Si} \approx 15$~pc, while at distances $R > R_{\rm Fe} \approx 14$~pc, the iron ions FeIV, FeV, FeVI follow the neutral atom FeI in abundance. Note that the results for Mg, Si and Fe ions depend on the assumptions made in our model: the atoms in the gas phase are in the neutral state at the initial time moment, and the atoms are released into the gas phase also in the neutral state during the thermal sublimation of dust grains.

Figure~\ref{fig5}c shows the abundances of S ions as a function of distance. The ionization potential of SI atom is 10.4~eV. The main ionization channel of the neutral SI atom is the photoionization by UV photons (in the cloud region where the UV radiation is not shielded). The transition SII/SI is located at a distance of $R_{\rm S} \approx 13-14$~pc (Fig.~\ref{fig5}). 
The main ionization channel of SII, and also of SI at $R > R_{\rm S}$, is the photoionization by photons in the X-ray wavelength range, which is accompanied by the emission of Auger electrons. The next most abundant sulfur ion after SI is SV at distances $R > R_{\rm S}$.
The number densities of SIII and SIV ions are low at all distances. The SIII and SIV number densities have close values at $R > R_{\rm S}$ -- the probabilities of the emission of one and two Auger electrons as a result of the photoionization of electron K-shell of SI have close values, about 0.08 \autocite{Kaastra1993}.

\begin{figure*}[h]
\centering
\includegraphics[width = 0.9\textwidth]{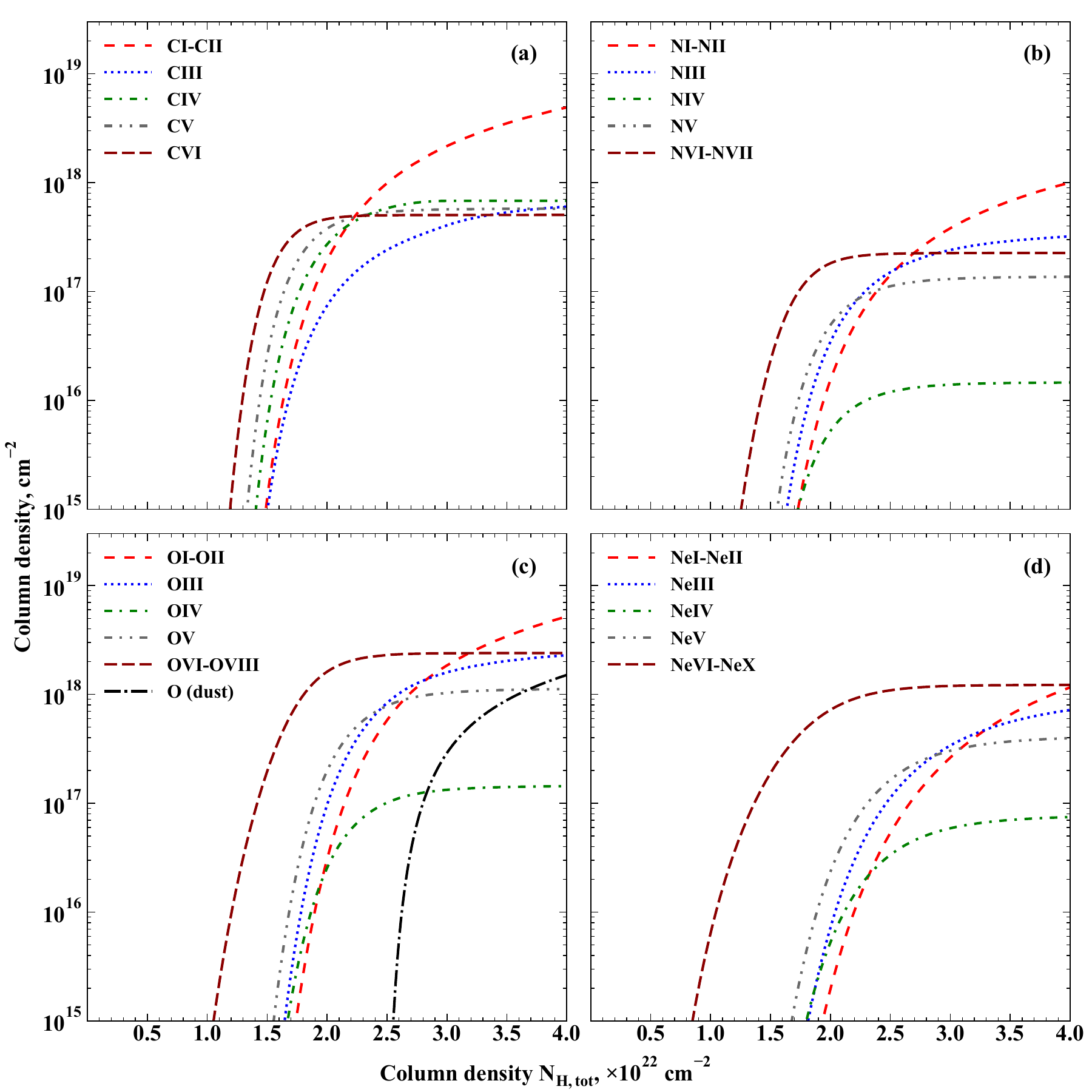}
\caption{\rm The column densities of C, N, O and Ne ions in various states of ionization versus hydrogen column density $N_{\rm H,tot}$. The column densities of completely stripped ions (CVII, NVIII, OIX, NeXI) are not shown in the graphs. The results correspond to the time $t_{\rm max} = 10^5$~s after the gamma-ray burst onset.}
\label{fig6}
\end{figure*}

\begin{figure*}[h]
\centering
\includegraphics[width = 0.9\textwidth]{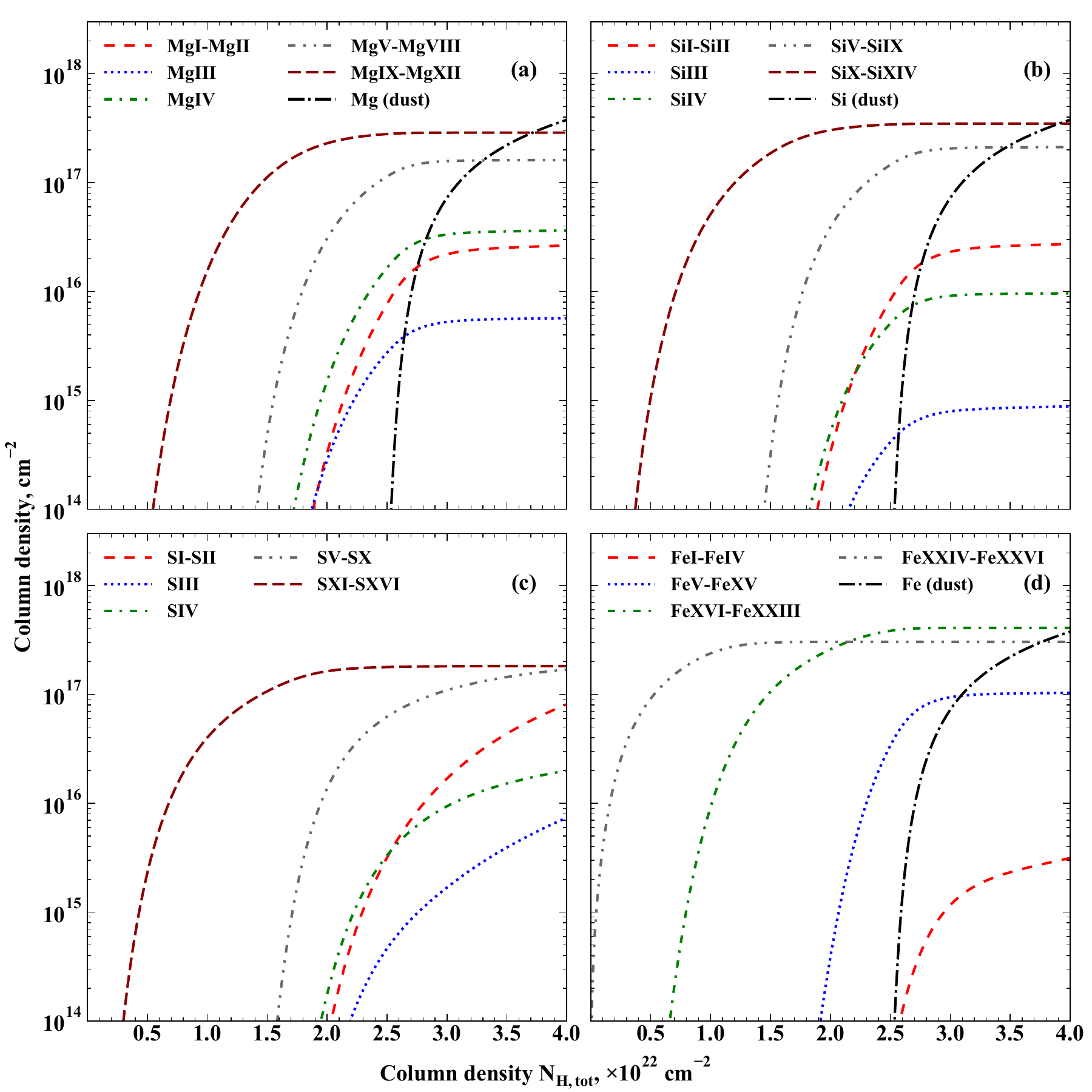}
\caption{\rm Same as in Fig.~\ref{fig6}, only for Mg, Si, S and Fe ions.}
\label{fig7}
\end{figure*}

\subsection{The column densities of metal ions}
Let us define the column density of ions of species $A$ at distance $R_j$ (the inner radius of the cloud shell $j$):

\begin{equation}
N_A(R_j) = \sum_{i = 1}^{j - 1} n_{A}^i (t_r) \Delta R,
\end{equation}

\noindent
where $n_{A}^i$ is number density of ions $A$ in the cloud shell $i$ at the same retarded time $t_r$. The column densities of ions can be measured using the observations of absorption lines of metal ions in the afterglow of the gamma-ray burst. In this section we present the results of our calculations of the column densities of metal ions as a function of the total column density of hydrogen nuclei, $N_{\rm H,tot} = n_{\rm H,tot} (R - R_{\rm min})$, where $R$ is the distance from the gamma-ray burst progenitor.
Figures~\ref{fig6} and \ref{fig7} provide our calculations at the retarded time $t_{\rm max} = 10^5$~s. The results for $N_{\rm H,tot} \leq 4 \times 10^{22}$~cm$^{-2}$ are given, corresponding to distances $R \leq 14$~pc at the gas density in question.

At hydrogen column densities $N_{\rm H,tot} < 2 \times 10^{22}$~cm$^{-2}$, the carbon ions are predominantly in the states of a high ionization CVI and CVII (Fig.~\ref{fig6}a). At higher hydrogen column densities, carbon is predominantly in the form of CI and CII ions (integrated along the line of sight). 
The upper boundary of the hydrogen column densities $N_{\rm H,tot}$, at which the column density of ions in the ionization state III or higher exceeds the column density of ions in the states I and II, increases with atomic number of the ion (Fig.~\ref{fig6}). For carbon this value of $N_{\rm H,tot}$ is about $2 \times 10^{22}$~cm$^{-2}$. And for neon, the column density of NeI--NeII is higher than the column density of NeIII--NeX only at $N_{\rm H,tot} > 4 \times 10^{22}$~cm$^{-2}$. 

The column densities of Mg, Si, and Fe ions in the state of ionization I--IV are much lower than the column densities of ions in the ionization state V or higher at all hydrogen column densities $N_{\rm H,tot}$, see Figs.~\ref{fig7}a, \ref{fig7}b, \ref{fig7}d. Thus, Mg, Si, and Fe in the gas phase are predominantly in the form of ions with a high charge (integrated along the line of sight).
At hydrogen column densities $N_{\rm H,tot} \gtrsim 3 \times 10^{22}$~cm$^{-2}$, the fraction of metal atoms Mg, Si, and Fe that are depleted into dust ceases to be negligible. The fraction of sulphur ions with a low charge (I--II) increases with the increase of the hydrogen column density $N_{\rm H,tot}$ (Fig.~\ref{fig7}c)
At $N_{\rm H,tot} \gtrsim 6 \times 10^{22}$~cm$^{-2}$, the column density of SI--SII ions is higher than the column density of sulphur ions with a higher charge. At the same time, the column densities of S ions in states of ionization III and IV are low for all $N_{\rm H,tot}$, that is a consequence of the photoionization of sulphur by X-ray photons and the emission of Auger electrons.

\begin{figure*}[h]
\centering
\includegraphics[width = 0.6\textwidth]{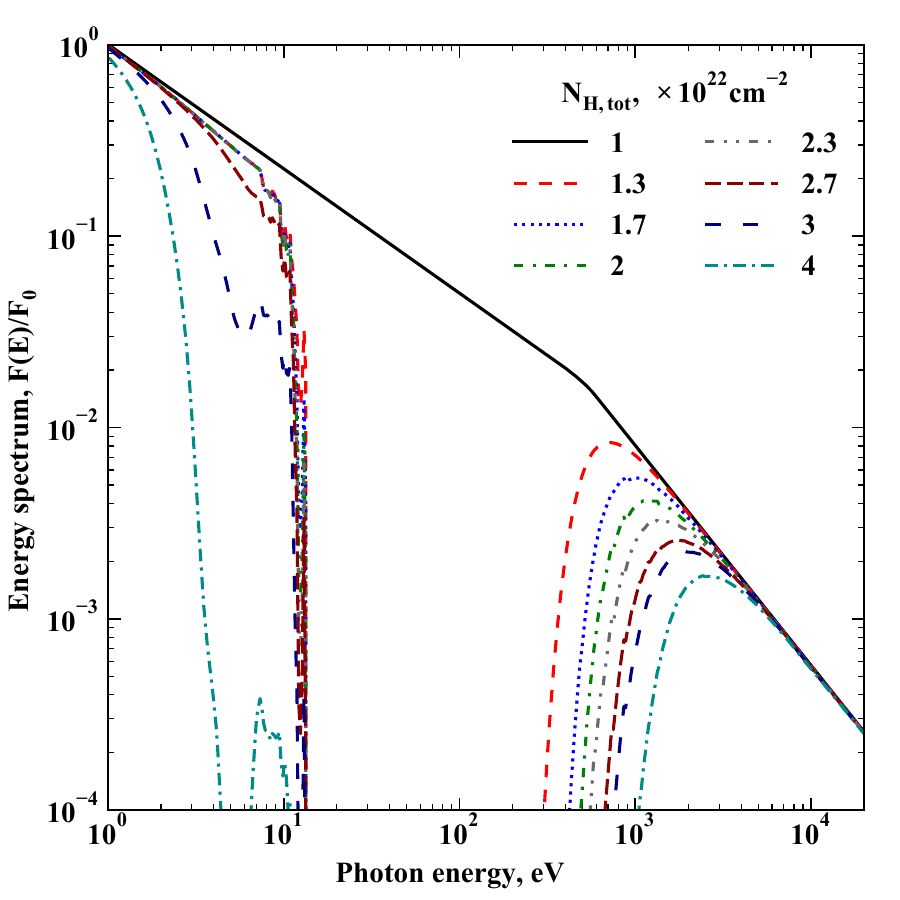}
\caption{\rm The energy spectrum of the gamma-ray burst afterglow after its passage through a cloud layer with a column density of hydrogen nuclei $N_{\rm H,tot}$. The spectrum is normalized to the value of the spectrum at photon energy $E_0 = 1$~eV. The results of the calculations are presented for different values of $N_{\rm H,tot}$. The results correspond to $t_{\rm max} = 10^5$~s.}
\label{fig8}
\end{figure*}

\subsection{The absorption of the gamma-ray burst afterglow in the gas--dust cloud}
Figure~\ref{fig8} presents the calculations of the energy spectrum of the afterglow at $t_{\rm max} = 10^5$~s after its passage through a cloud layer of various thicknesses. The column density $N_0$ of the fully ionized gas layer is $\approx 10^{22}$~cm$^{-2}$. This gas has almost no effect on the observed afterglow spectrum (the effect from the Compton scattering by electrons and the photoabsorption of radiation by metal ions in ultra-high ionization states is negligible).
The radiation in the UV wavelength range with a photon energy above 13.6~eV is completely absorbed by H, H$_2$, He in a partially ionized cloud layer for all $N_{\rm H,tot} > N_0$. The dust-destruction radius $R_{\rm d} \approx 9-10$~pc corresponds to the hydrogen column density $N_{\rm d} \approx 2.7 \times 10^{22}$~cm$^{-2}$. At hydrogen column densities $N_{\rm H,tot} < N_{\rm d}$, the radiation in the optical and UV wavelength ranges ($h\nu < 13.6$~eV) encounters weak absorption.
In the X-ray wavelength range, the absorption of radiation occurs for all values of the parameter $N_{\rm H,tot} > N_0$. At $N_{\rm H,tot} \leq N_{\rm d}$ and $1 \lesssim h\nu \lesssim 5$~keV, the metal ions make a major contribution to the radiation absorption (for solar metallicity). At higher photon energies, the Compton scattering by free electrons and the Compton ionization are the main photon absorption processes (but the optical depth of the cloud is small at such photon energies).
The upper energy threshold for the total absorption of radiation in the X-ray part of the energy spectrum gradually moves toward higher energies as the hydrogen column density $N_{\rm H,tot}$ increases. The metal ions are in various ionization states and, therefore, the absorption edges corresponding to the ionization of electron K- and L-shells of ions are weak and barely noticeable in the spectrum, see Fig.~\ref{fig8}. 

\section{Discussion}
\subsection{The ionization structure of the cloud}
The gamma-ray burst afterglow makes a major contribution to the fluence of the radiation in both UV and X-ray wavelength ranges. The contribution of the prompt phase of the gamma-ray burst to the fluence of the radiation in the X-ray wavelength range ($0.3-10$~keV) is about 20 per cent in our model (this roughly corresponds to the gamma-ray bursts' prompt efficiency $\eta$).

The UV and soft X-ray radiation ($h\nu > 13.6$~eV) is efficiently shielded by the gas--dust cloud (Fig.~\ref{fig8}) and, therefore, the X-ray radiation makes a major contribution to the ionization of H$_2$ molecules and metal ions at distances $R > R_{\rm ion}$. The model of the synchrotron radiation of the forward shock in the homogeneous circumburst medium predicts the following time dependence of the afterglow intensity in the X-ray range:

\begin{equation}
F_{\nu} \propto t^{-\alpha},
\end{equation}

\noindent
where $\alpha = -3(1-p)/4$ in the slow cooling regime ($\nu < \nu_{\rm c}$), and $\alpha = -(2-3p)/4$ in the fast cooling regime ($\nu > \nu_{\rm c}$), $p$ is the electron spectral index, $\nu_{\rm c}$ is the synchrotron frequency of an electron whose cooling time equals the dynamical time \autocite{Granot2002}.
In our model, $p = 2.3$, and $\alpha \approx 1$ and $\alpha \approx 1.2$ for slow and fast cooling. At $t = 10^3$~s, the cooling frequency equals $h \nu_{\rm c} \approx 4$~keV (Fig.~\ref{fig1}). The ionization rate of the gas by X-ray radiation decreases relatively slowly with time, which indicates the need to perform calculations for long time scales ($t_{\rm max} \gtrsim 10^4$~s).
According to our calculations, the gas ionization fraction at a distance of 15~pc from the burst progenitor increases from $1.5 \times 10^{-4}$ to $1.7 \times 10^{-4}$ in the time interval from $10^4$ to $10^5$~s (for solar metallicity). The distance from the gamma-ray burst progenitor to the ionization front also increases -- the increase in the radius of the ionization front is 0.4~pc (10 per cent) in the time interval from $10^4$ to $10^5$~s.

The increase in the gas ionization fraction stops when the rate of gas ionization becomes comparable to the rate of ion recombination. The rate of gas ionization at a distance of 15~pc at a time $10^5$~s is $\sim 10^{-10}$~s$^{-1}$. The characteristic rate of recombination of metal ions is $\alpha n_{e}$, where $\alpha \sim 10^{-12}-10^{-10}$~cm$^{3}$ s$^{-1}$ at $T_{e} \approx 10^3$~K \autocite{Pequignot1991}, while the electron number density equals $n_e = 0.2$~cm$^{-3}$ at a distance of 15~pc. Thus, for the time scales and distances being considered in this paper the rates of recombination reactions are low.

\subsection{The absorption of radiation in the X-ray wavelength range}
The observed values of the hydrogen column density $N_{\rm HX}$ that determines the absorption of radiation of the gamma-ray burst afterglow in the X-ray wavelength range lie in the range from low values to $\approx 10^{23}$~cm$^{-2}$, and the mean value is about $10^{22}$~cm$^{-2}$ \autocite{Campana2010,Campana2012}. In the model by \textcite{Watson2013} the radiation of the gamma-ray burst afterglow is absorbed in the HII region where the gamma-ray burst occurred. 
In their model the contribution of metal ions to the radiation absorption is small due to the ionization of metal ions to a high degree by the gamma-ray burst radiation, while the radiation is absorbed by helium ions. \textcite{Watson2013} considered the observed increase in the average ratio $N_{\rm HX}/A_{\rm V}$ with redshift $z$ as evidence that the radiation is absorbed by He ions rather than metal ions.

In our model the absorption of radiation takes place in a partially ionized molecular gas at distances larger than the ionization radius, $R > R_{\rm ion}$. At $N_{\rm H,tot} - N_0 = 10^{22}$~cm$^{-2}$ and photon energy 1~keV, the contribution of metal ions, HeI and H$_2$ to the optical depth is approximately 35, 35 and 25 per cent, respectively, in our model (for solar metallicity). The contribution of metal ions to the radiation absorption increases with photon energy, since the photoionization of inner electron shells of metal ions Si, S, Fe takes place at higher energies.
At photon energies $h\nu > 5-10$~keV, the Compton ionization of H$_2$ and He mainly contributes to the absorption of radiation. The contribution of HeII to the absorption is small. The optical depth due to the photoionization of electron K-shells of iron ions is small, $\sim 0.01$ (the ionization energies of electron K-shells of iron ions lie in the range $7-9$~keV).

\textcite{Cucchiara2015} studied the afterglow spectra of gamma-ray bursts in which ${\rm Ly}\alpha$ absorption line is observed. According to their results, the gas metallicity in the host galaxies of gamma-ray bursts at redshift $z = 3.5-6$ is, on average, about 10 per cent of the solar metallicity. On might expect that the contribution of HeI and H$_2$ to the absorption of afterglow radiation in the X-ray range increases with decreasing metallicity.
According to our calculations for $\rm [M/H] = -1$, the contribution of He atoms and H$_2$ molecules to the radiation absorption is dominant: the contribution of HeI and H$_2$ to the optical depth at photon energy 1~keV is about 55 and 40 per cent, respectively. The results of our calculations confirm the suggestion made by \textcite{Watson2013} that the photoionization of He atoms plays an important role in the absorption of radiation in the X-ray wavelength range.
In our model the absorption of radiation takes place in a dense molecular cloud near the gamma-ray burst progenitor. \textcite{Starling2013} also showed that in the case of a cold neutral gas the contribution of H/H$_2$ and HeI to the absorption of radiation in the X-ray range is significant even for solar metallicity.

The column density of atomic hydrogen $N_{\rm HI}$ is proportional to the number density of H atoms in the cloud at the initial time, i.e. is a model-dependent parameter. In our model $n_{\rm H} = 0$ at $t = 0$, and, therefore, $N_{\rm HI}$ is low and does not depend on $N_{\rm H,tot}$. After the passage of the gamma-ray burst radiation through the cloud, the atomic hydrogen is predominantly near the ionization front, and $N_{\rm HI} = 1.5 \times 10^{20}$~cm$^{-2}$ (Fig.~\ref{fig3}).  
According to our calculations, the column densities of Mg, Si, and Fe ions in states of ionization I--IV are much lower than the column densities of ions with a higher charge for all of the values of $N_{\rm H,tot}$ in question, see Fig.~\ref{fig7}. These results are consistent with observational data -- the analysis of the time variability of the absorption line profiles of metal ions with a low charge implies that the absorption in the lines of these ions occurs at large distances from the gamma-ray burst progenitor \autocite{Vreeswijk2007}.  
Furthermore, according to our calculations, there is no absorption of optical radiation by dust in our model at $N_{\rm H,tot} < N_{\rm d} \approx 2.7 \times 10^{22}$~cm$^{-2}$, where $N_{\rm d}$ is the hydrogen column density that corresponds to the dust-destruction radius. Our model predicts the presence of H$_2$ absorption lines in the gamma-ray burst afterglow spectrum. The absorption lines of H$_2$ molecule are indeed observed in the afterglow spectra of some gamma-ray bursts \autocite{Bolmer2019}. 

\section{Conclusions}
A model has been constructed for the passage of a gamma-ray burst through a dense molecular cloud located near the burst progenitor. The gas density in the model is taken equal to $n_{\rm H,tot} = 10^3$~cm$^{-3}$. The intensity and duration of the gamma-ray burst radiation in the UV wavelength range determine the radius of the ionization front and the dust-destruction radius. The size of the boundary between the fully ionized gas and the neutral gas is about 0.05~pc; at this distance the gas ionization fraction drops by a factor of $\approx 10$.

The ionization of metal ions by X-ray radiation determines the gas ionization fraction in the cloud region where the gas is predominantly neutral. The metal ions are ionized via the photoionization of their inner electron shells which is accompanied by the emission of Auger electrons. This gives rise to ions in a high ionization state.
The column densities of Mg, Si, and Fe ions in states of ionization I--IV are much lower than the column densities of ions in the state of ionization V and higher. The column density of sulphur ions SI and SII becomes higher than the column density of sulphur ions in higher ionization states only for high hydrogen column densities in the cloud, $N_{\rm H,tot} > 6 \times 10^{22}$~cm$^{-2}$, while the column density of sulphur ions SIII and SIV is low for all $N_{\rm H,tot}$. The photoionization of metal ions by UV radiation takes place only at distances smaller than the dust-destruction radius and for ions with an ionization threshold below 13.6~eV.  

The results of our calculations confirm the assumption by \textcite{Watson2013} that the photoionization of He atoms plays an important role in the absorption of radiation in the X-ray wavelength range. For a low metallicity, $\rm [M/H] \leq -1$, the role of He  atoms is dominant. The numerical model constructed in this paper will be used in future to fit the gamma-ray burst afterglow spectra in the X-ray wavelength range and to interpret the observational data for the absorption lines of metal ions. 

This work was supported by RSF grant no. 21-12-00250.


\printbibliography 

\end{document}